\documentclass[prl,aps,showpacs,twocolumn,superscriptaddress,floatfix]{revtex4-1}

\usepackage{epsfig}
\usepackage{amsmath}
\usepackage{amssymb}
\usepackage{amsfonts}
\usepackage{mathptmx}
\usepackage{eucal}
\usepackage{bm}
\usepackage{color}
\usepackage[colorlinks,linkcolor=blue,citecolor=blue]{hyperref}

\usepackage{epstopdf}

\usepackage{graphicx}

\usepackage{natbib}

\begin{document}
\title{Quantum interference hybrid spin-current injector}
\author{F. Giazotto}
\email{giazotto@sns.it}
\affiliation{NEST, Instituto Nanoscienze-CNR and Scuola Normale Superiore, I-56127 Pisa, Italy}
\author{F. S. Bergeret}
\email{sebastian\_bergeret@ehu.es}
\affiliation{Centro de F\'{i}sica de Materiales (CFM-MPC), Centro
Mixto CSIC-UPV/EHU, Manuel de Lardizabal 4, E-20018 San
Sebasti\'{a}n, Spain}
\affiliation{Donostia International Physics Center (DIPC), Manuel
de Lardizabal 5, E-20018 San Sebasti\'{a}n, Spain}
\affiliation{Institut f\"ur Physik, Carl von Ossietzky Universit\"at, D-26111 Oldenburg, Germany}

\begin{abstract}
We propose a quantum interference spin-injector nanodevice consisting of a superconductor-normal metal hybrid loop connected to a superconductor-ferromagnet bilayer via a tunneling junction.  We show that for certain  values of  the applied  voltage bias across the tunnel barrier and the  magnetic flux through the loop the spin-current can be fully polarized. Moreover, by tuning the magnetic flux one can switch the sign of the spin polarization. This operation can be performed  at frequencies within the tens of GHz range. We explore the nanodevice in a wide range of parameters,  establish the optimum conditions for  its experimental realization and discuss its possible applications. 
\end{abstract}
\pacs{72.25.-b,85.75.-d,74.50.+r}

\maketitle

  {\it INTRODUCTION.-} Generation of strongly   spin-polarized currents and their control in nanoscale circuits  is  highly desirable  in the field of \emph{spintronics} \cite{Zutic}.  
  In this context, there have been a number of proposals to achieve highly spin-polarized currents using different nanodevices \cite{Stefanski,Csonka,Feng,Zozoulenko,Chen,Giazotto2003}.
  Such a spin-currents can   produce  a dynamical  switching by means of spin-transfer torque \cite{Slonczewski,Berger}
  of the magnetization in multilayer ferromagnetic structures\cite{Buhrman}.  Such a switching procedure is used in  magnetic random access memories,   where the magnetic configuration is controlled by spin-polarized currents.   
 Usually, the source of spin-polarized currents is a ferromagnet  with highly-polarized conduction electrons.
 Therefore, materials with half-metallic behavior, as for example CrO$_2$, are ideal candidates for spin-injectors. However, their growth  in hybrid nano circuits still remains  a challenge. 
 Alternatively,   strongly spin-polarized currents can be created in  hybrid structures consisting of  a superconductor (S) and a ferromagnetic (F)  layer \cite{Nazarov2002,Giazotto2008}
  tunnel-coupled to a normal  metal (N).  It was shown that  in such hybrid structures fully spin-polarized currents can be induced. 
  In particular,  the NISF structure (I denotes an insulating layer) studied in Ref. \cite{Giazotto2008} allows for
  tuning of the sign and magnitude of the  spin polarization by changing the  bias voltage.

  Here, we propose a spintronic nanodevice that,  on the one hand,  is able to provide  strongly spin-polarized currents, and on the other hand 
  allows  for a switching  of the current polarization not only by means of a voltage bias  but also by an external magnetic flux.  The switching time between
   positive and negative current polarization can be achieved in the  nanosecond range or faster.
   The device consists of a superconducting loop (S$_1$) interrupted by a N wire of length $L$. In addition, a
   SF bilayer is tunnel-coupled to N through a junction with normal-state resistance $R_t$ [see Fig. 1(a)].
    We assume  (i) a good contact between the S and F layers and (ii) that $R_t$ is much larger than the SF contact resistance which ensures  the bilayer to be in local equilibrium.
    $t_S$ ($t_F$) labels the S (F) layer thickness,
		and the SF bilayer is kept at a constant voltage $V$, while the other end of the structure is grounded.
    Except for  the F layer, the setup shown in Fig. 1(a) resembles the ones investigated in recent 
    experiments on hybrid  nanostructures \cite{Giazotto2010,Meschke}.  
\begin{figure}[tb]
\includegraphics[width=\columnwidth]{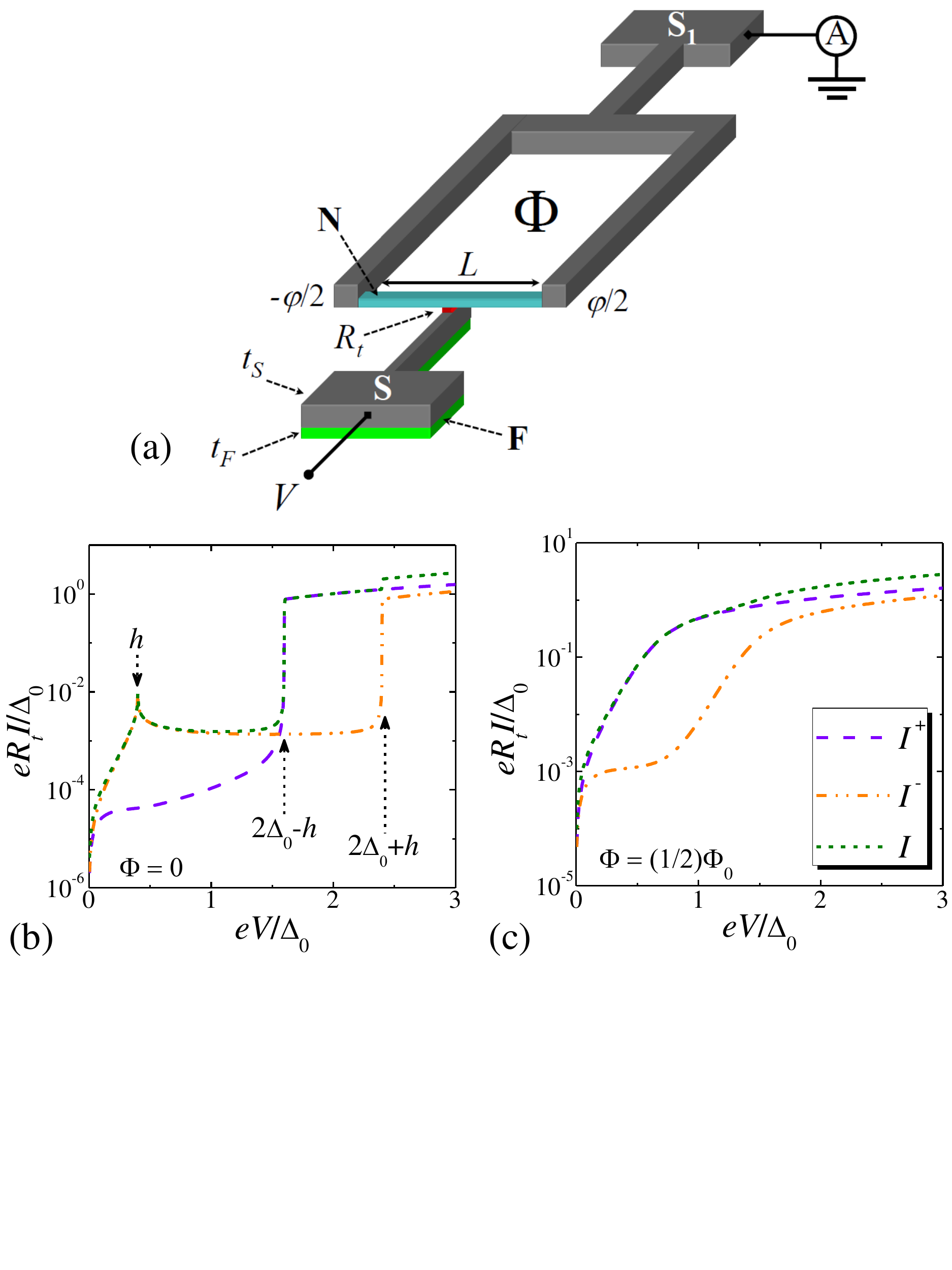}\vspace{-3mm}
\caption{(a) Scheme  of the proposed quantum interference spin-current injector. $\varphi$ is the quantum phase difference in S$_1$ whereas $\Phi$ is the externally applied magnetic flux (see text). 
(b)  Current vs voltage characteristics for spin-up ($I^+$) species, spin-down ($I^-$) species and their sum ($I$) for zero applied magnetic flux. (c) The same as in panel (b) but for a half-integer flux quantum applied through the loop. In (b) and (c) we set $h=0.4\Delta_0$ and $T=0.2T_c$, where $\Delta_0$ is the zero-field and zero-temperature energy gap, whereas $T_c$ is the superconducting critical temperature.}
\label{fig1}
\end{figure}
     
    Our hybrid interference spin-current injector operates as follows: 
     by applying a voltage bias a finite quasiparticle current flows through  the structure. 
	The amplitude of the resulting current depends on the density of states (DoS) of both  the 
     SF bilayer and the N wire. The former shows a Zeeman-splitting induced by the exchange field of the F layer, whereas the latter is modified by the \emph{proximity} effect induced from the nearby contacts with the S$_1$ loop.  
     As it is shown below, the current flowing through the structure can be strongly spin-polarized. 
		In addition to the  dc voltage bias, the device can also  be operated by an externally-applied magnetic flux. 
		By neglecting the loop inductance, the magnetic flux $\Phi$  fixes the superconducting phase difference across the SN boundaries  
    according to $\varphi=2\pi\Phi/\Phi_0$, where $\Phi_0$ is the flux quantum.   
		Since the DoS in the N wire depends on the phase difference \cite{Belzig},  by varying  the magnetic flux through the loop one can modify 
    the DoS in N and,  in turn,  the electric current and its spin-polarization.  Therefore, our device can be used as a \emph{phase-tunable} spin-injector. One of the advantages of our setup with respect to a voltage tunable spin-injector based on the results of Refs. \cite{Nazarov2002,Giazotto2008} is that the switching speed provided by the magnetic flux allows, in principle, for a much higher operation frequency.

{\it THE MODEL.-}In order to model the spin-current injector we consider a SF system which may consists either of two  thin S and F layers  in good electric contact, or of  a superconducting layer in contact with a ferromagnetic insulator. 
In particular, we consider the situation where  the thickness of the S layer is smaller than the superconducting coherence length, and the F thickness is smaller than the length of the condensate penetration into the ferromagnet. 
In such a case the ferromagnet induces a homogeneous effective exchange field ($h$) in S through proximity effect which modifies the superconducting gap ($\Delta_0$). $h$ and the effective gap in S ($\Delta$) are given by $h/h_0=\nu_Ft_F(\nu_St_S+\nu_Ft_F)^{-1}$ and $\Delta/\Delta_0=\nu_St_S(\nu_St_S+\nu_Ft_F)^{-1}$, respectively, where $h_0$ is the original exchange field existing in the ferromagnetic layer and
$\nu_S$ ($\nu_F$) is the normal-state DoS in S (F) at the Fermi energy.
If $\nu_S=\nu_F$ and for $t_F\ll t_S$ it follows that $\Delta\approx \Delta_0$ while $h/h_0\approx t_F/t_S\ll 1$. 
 The  effect of $h$ on the superconductor  leads to a spin-dependent BCS-like DoS shifted by the effective exchange energy (similarly to what happens for a Zeeman-split superconductor in a magnetic field) \cite{Bergeret2001, Bergeret2012,Meservey1,Meservey2}. 
The total DoS of the SF layer is then given by the sum of the spin-up ($\nu_+$) and spin-down ($\nu_-$)  density of states which can be written as \cite{Meservey1,Meservey2}
\begin{equation}
\nu_{\pm}(E)=\frac{1}{2}\left|{\rm Re}\left[\frac{E\pm h+i\Gamma}{\sqrt{(E\pm h+i\Gamma)^2-\Delta^2(h,T)}}\right]\right|\; ,\label{DosSF}
\end{equation}
where $E$ is the energy measured from the condensate chemical potential, and $T$ is the temperature.
The effective superconducting order parameter, $\Delta(h,T)$, depends on both the temperature and the magnitude of the exchange field, and
has to be determined self-consistently 
from the gap equation 
$\text{ln}(\Delta_0/\Delta)=\int_0^{\hbar\omega _D}d\varepsilon (\varepsilon ^2+\Delta^2)^{-1/2}[f_+(\varepsilon)+f_-(\varepsilon)]$, where $f_{\pm}(\varepsilon)=\left\{1+\text{exp}[\frac{1}{k_B T}(\sqrt{\varepsilon ^2+\Delta^2}\mp h)]\right\}^{-1}$, $\Delta_0=1.764 k_B T_c$ is the zero-temperature order parameter in the absence of exchange field, $T_c$ is the superconducting critical temperature, $k_B$ is the Boltzmann constant, and $\omega_D$ is the Debye frequency.
The parameter $\Gamma$  in Eq. (\ref{DosSF}) accounts for  the inelastic scattering energy rate within the relaxation time approximation \cite{Dynes1984,Pekola2004}.  
Tunneling conductance measurements on SIS junctions \cite{Moodera1990}, where I is a ferromagnetic insulator, have shown  the accuracy of the  description of the DoS provided by Eq. (\ref{DosSF}). 

We assume that the  tunnel junction between the SF bilayer and the normal metal  is situated  in the middle of the N wire, and that the  resistance of the tunneling contact $R_t$ is much larger than  the normal-state resistance $R_N$ of the N wire and the resistance   $R_{SN}$ of the  SN interfaces. Therefore, the voltage drop occurs entirely at the SF/N contact. 
Moreover, we assume  the wire transverse dimensions to be much smaller than $L$, so that it can be considered as quasi-one-dimensional (1D), and 
neglect any spatial later extension of the tunnel junction \cite{footnote}.  For the sake of clarity in our analysis
we choose identical superconductors S and S$_1$ for  the nanodevice of Fig 1(a).  

The total quasiparticle current is  given by the sum of the spin-up and spin-down contributions,  $I=I^++I^-$ where 
\begin{eqnarray}
I^\pm=\frac{1}{2eR_t}\int^{+\infty}_{-\infty} &dE& \nu_{\pm}(E-eV)\nu_N(E,\Phi)\times\nonumber\\
&&[\tanh ({E-eV}/2k_BT)-\tanh(E/{2k_BT})],\,\,\,\,\,\,\,\,\,\,\,\,\label{current}
\end{eqnarray}
 $\nu_N(E,\Phi)$ is the DoS in the middle  of the N wire, and $e$ is the electron charge.  
 The exact form of $\nu_N(E,\Phi)$ can be obtained from the knowledge of the retarded  quasiclassical Green's function. 
 The latter  is the solution 
of   the 1D Usadel equation in the N region \cite{usadel} 
 \begin{equation}
 \partial_x\left(\hat g\partial_x \hat g\right)+i\frac{E}{E_{Th}}\left[\tau_3,\hat g\right]=0,\label{Usadel}
 \end{equation}
 where $\hat g$ is a 2$\times$2 matrix in the Nambu space,   $E_{th}=\hbar D/L^2$ is the Thouless energy and $D$ is the diffusion constant in N.
 All lengths are given in units of $L$. 
The DoS in the normal wire is then determined by the real part of the (1,1) component of $\hat g$.  
It is known that due to the proximity effect $\nu_N (E,\Phi)$ shows 
a minigap whose size depends, among other parameters,  on the phase difference across  the N wire\cite{Giazotto2010,Meschke,Cuevas2007,Saclay}.
 Thus, by varying  the magnetic flux through the loop one can control the size of the minigap in N, which is maximized for $\Phi=0$ and vanishes for $\Phi=(1/2)\Phi_0\textrm{mod}[2n\pi]$, where $n$ is an integer. 
 
Equation (\ref{Usadel}) is supplemented by boundary conditions describing the transmissivity of the SN interfaces \cite{KL}
 \begin{equation}
 \hat g\partial_x\hat g=\pm\frac{\gamma}{2}\left[\hat g,\hat g_{R(L)}\right]\; ,\label{BC}
 \end{equation} 
 where $\gamma=R_N/R_{SN}$, and $\hat g_{R(L)}$ are the bulk  BCS Green's functions.  
For simplicity we have assumed that the normal-state conductivity of the S and N parts of the wires are the same.  
We notice that in the case of a perfectly-transmissive SN interface, $R_{SN}\rightarrow 0$, and Eq. (\ref{BC}) imposes the continuity of $\hat g$ at the boundary.  
Furthermore, we neglect the suppression of the order parameter in  S$_1$ at the NS$_1$ boundaries due to \emph{inverse} proximity effect \cite{Saclay,cuevas}. We expect however this effect to be very small in a real nanostructure by making the cross section of the loop much larger than that of the N wire.
Finally, in all the following calculations we set $\Gamma =10^{-4}\Delta_0$ as a representative value describing realistic tunnel junctions \cite{Pekola2004}.
\begin{figure}[tb]
\includegraphics[width=\columnwidth]{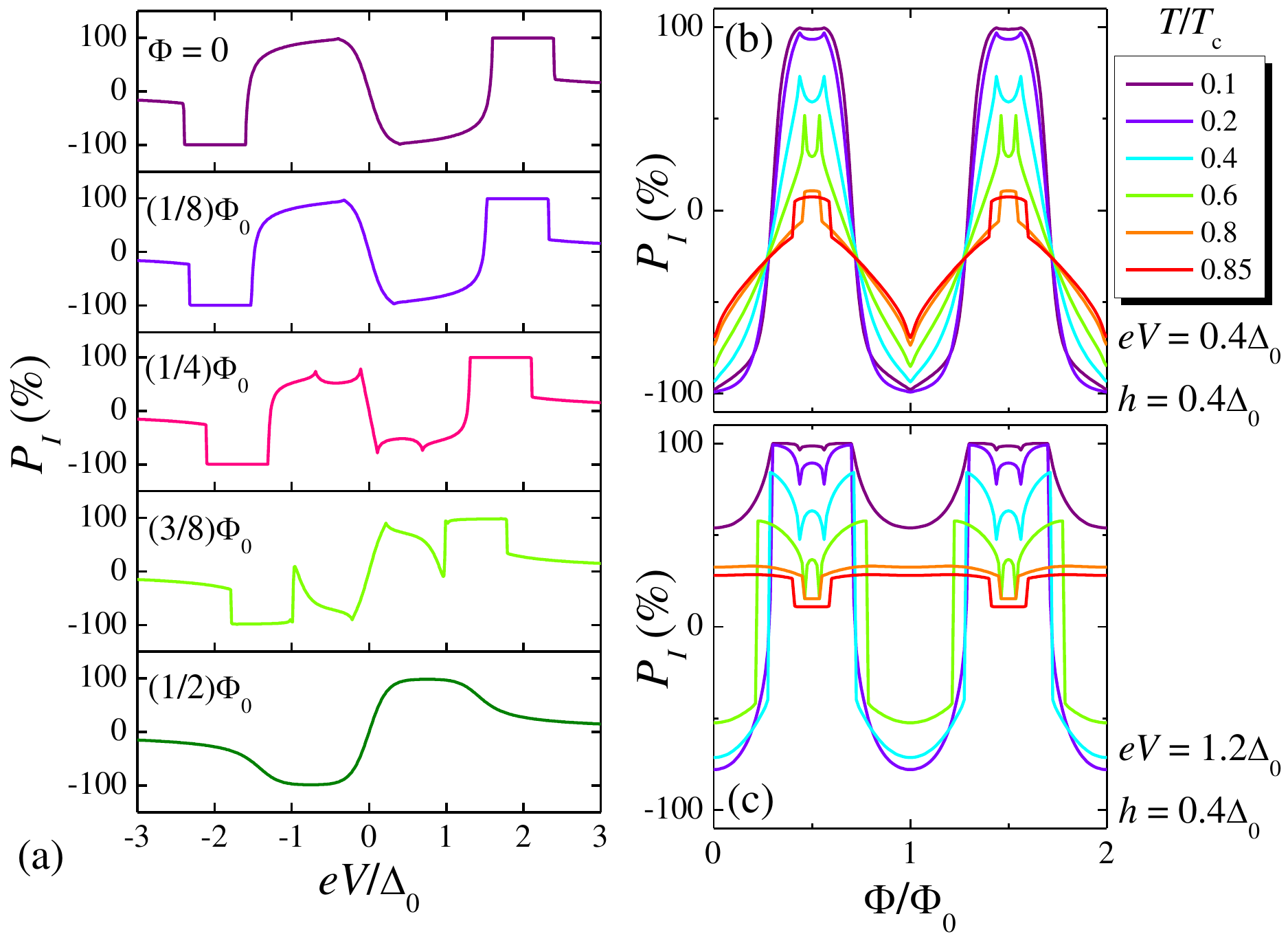}\vspace{-2mm}
\caption{(a) Voltage dependence of the current spin-polarization $P_I$ calculated for different magnetic flux values at $T=0.2T_c$. 
Flux dependence of the current polarization calculated for different temperatures at $eV=0.4\Delta_0$, (b), and $eV=1.2\Delta_0$, (c). In all calculations we set $h=0.4\Delta_0$.}
\label{fig2}
\end{figure}

{\it RESULTS AND DISCUSSION.-} We   first  consider the  case of a  \emph{short} N bridge satisfying the condition $E_{Th}\gg\Delta$, and assume perfectly-transmissive SN contacts.
In this regime the spin-current injector performance is optimized since the proximity effect in N  is maximized.
In this case the DoS  in the middle of the wire can be obtained analytically,  \cite{Heikkila2002}
  \begin{equation}
 \nu_{N}(E,\Phi)={\rm Re}\left|\left[\frac{E+i\Gamma}{\sqrt{(E+i\Gamma)^2-\Delta^2(T)\cos^2(\pi \Phi/\Phi_0)}}\right]\right|.\label{Dos0}
\end{equation}
It has a  BCS-like form with an effective gap, $\Delta_g=\Delta(T)|\cos(\pi \Phi/\Phi_0)|$,
 whose magnitude depends on the magnetic flux $\Phi$.   In particular for $\Phi=(1/2)\Phi_0$ the minigap is completely closed.
Substituting Eqs. (\ref{DosSF},\ref{Dos0}) into  the expression for the current,  Eq. (\ref{current}),  we compute the spin-currents  $I^\pm$. 
The voltage dependence of  $I^\pm$  and of the total current through the device are  shown in  Figs. 1(b) and 1(c) for zero and half-integer flux quantum, respectively.  Within  certain ranges of voltage, the spin-up and spin-down currents can differ by several orders of magnitude.
In  the  zero  flux case  [see Fig. 1(b)],  the spin-down current dominates the transport ($I^-\gg I^+$) 
if $eV<2\Delta_0-h$, whereas for $ eV> 2\Delta_0-h$ the opposite occurs ($I^+\gg I^-$).
 The thresholds for the onset of large quasiparticle current [see panel (b)] correspond to the sum of the gaps on both sides of the barrier ($eV=2\Delta_0\pm h$),  i.e.,  in the SF and in the  N layer. This is  in analogy to  the quasiparticle \emph{I-V}  characteristic of a conventional SIS tunnel junction \cite{Tinkham}.   
We note that a Josephson supercurrent, although pretty small,  can flow through the device\cite{Giazotto2010,Meschke,Giazotto2011}, however in our case a spin-polarized current is achieved only if a  finite bias voltage (i.e., $V\neq 0$) is applied. 
For half-integer flux quantum values [see Fig. 1(c)] the behavior is modified, i.e., $I^+$ is substantially larger than $I^-$ in the whole range of voltage. 
As a consequence, it turns out that by applying a magnetic flux through the loop the voltage dependence of the  current polarization, defined as
\begin{equation}
 P_I(\Phi,V)=\frac{I^+-I^-}{I^++I^-},
\end{equation}
 may change drastically.  
This is shown explicitly in Fig. 2(a) where the $P_I(V)$ dependence is plotted for different values of the applied flux at $T=0.2T_c$ and $h=0.4\Delta_0$. 
We emphasize that high spin polarization of both signs (i.e., up to $\sim \pm 100\%$) can be achieved in the nanostructure within suitable voltage bias windows and magnetic flux. 
\begin{figure}[tb]
\includegraphics[width=\columnwidth]{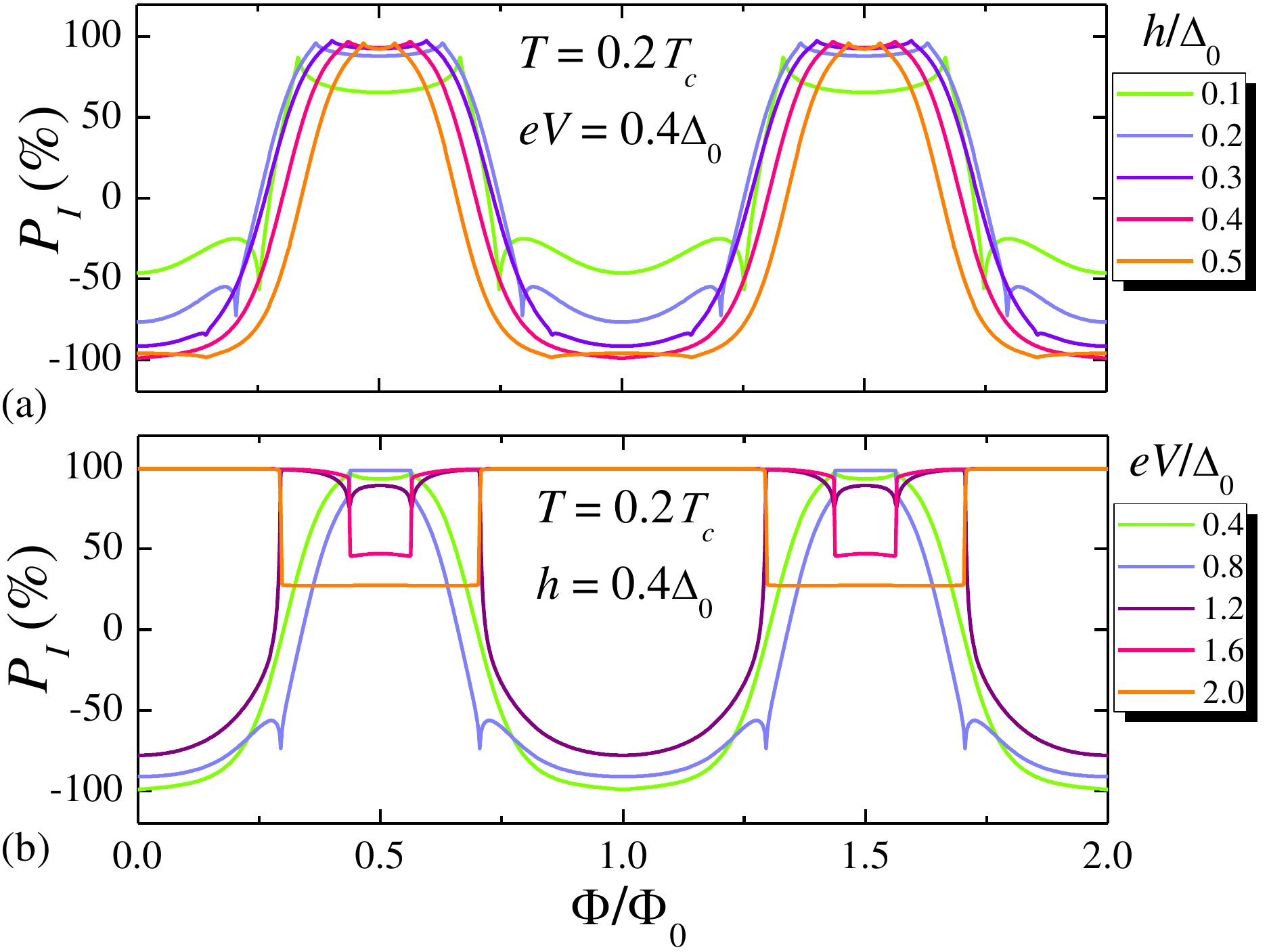}\vspace{-2mm}
\caption{(a) $P_I(\Phi)$ dependence calculated for different values of the effective exchange field in the SF bilayer at $T=0.2T_c$ and $eV=0.4\Delta_0$.
(b) $P_I(\Phi)$ dependence calculated for different values of the applied voltage at $T=0.2T_c$ and $h=0.4\Delta_0$.}
\label{fig3}
\end{figure}

We demonstrate in this way that,  in addition to the voltage dependent switching of the current-polarization, our device is able to switch the magnitude and sign of $P_I$ by tuning the magnetic flux through the loop. 
In Figs. 2(b) and 2(c) we show the $P_I(\Phi)$ dependence for two different values of the voltage bias at different temperatures.  The maximum values of $P_I$ are typically achieved at low temperatures, $T\leq0.2 T_c$, where the current is almost full-polarized. We stress, however, that even for temperatures up to $\sim0.6T_c$ a sizeable polarization switching effect is still observable. 

In Fig. 3 we show the flux dependence of the current spin polarization calculated at $T=0.2 T_c$ for different values of the effective exchange field in the SF structure [panel (a)], and for different voltage bias applied across the junction [panel (b)]. 
From these figures it follows that  strong  spin-polarized currents and sign switching (i.e., around $\sim \pm 100\%$)  can be achieved for large  enough values of $h$ and subgap voltages. Large values of $P_I$ can also be achieved for $|eV|>1.2\Delta_0$, though the sign  switching is not possible for such large bias [see Fig. 3(b)].

All the results presented above have been obtained in the limit of  a short N wire, i.e.,  when  $E_{Th}\gg \Delta_0$, and for perfectly-transmitting SN interfaces. 
In the case of an arbitrary Thouless energy (i.e., for arbitrary wire \emph{length}) and arbitrary transparency of the SN interfaces 
we have  solved numerically  Eqs. (\ref{Usadel}-\ref{BC}) in the N region 
for the retarded Green's function to  obtain the DoS in the middle of the wire, and computed the currents $I^\pm$ from Eq. (\ref{current}). 
In Fig. 4 we show the results for the  dependence of the current spin-polarization on the magnetic flux. 
Panel (a)  shows this dependency for different Thouless energies and  a  highly-transparent SN interface ($R_{SN}\rightarrow 0$) assuming $eV=h=0.4\Delta_0$.
One can see that by increasing the length $L$, {\it i.e.,} by decreasing $E_{Th}$, 
the range of switching is suppressed. 
Nevertheless,  a large spin polarization modulation amplitude is still  present up to $E_{Th}\sim0.5 \Delta_0$. By choosing, for instance, aluminum (Al) with $\Delta_0=200\mu$eV and silver (Ag) with $D=0.02$m$^2$s$^{-1}$ as prototypical materials to implement the spin-current injector,  
this value would correspond to  a realistic length of N wire, $L=\sqrt{\hbar D/E_{Th}}\simeq 360$ nm. 
It is easily understandable that  if the wire is very long, i.e., for $E_{Th}\ll\Delta_0$, the proximity effect in the middle of the N region is strongly weakened with the consequences that the DoS is  almost magnetic-flux independent,  
and  the spin polarization  does not show any significant changes as a function of $\Phi$. 
We stress, however,  that even in the case of a long wire the achievable current polarization may be strong  providing  a proper choice of the bias voltage  [see, for example, the blue curve in Fig. 4(a)].

\begin{figure}[tb]
\includegraphics[width=\columnwidth]{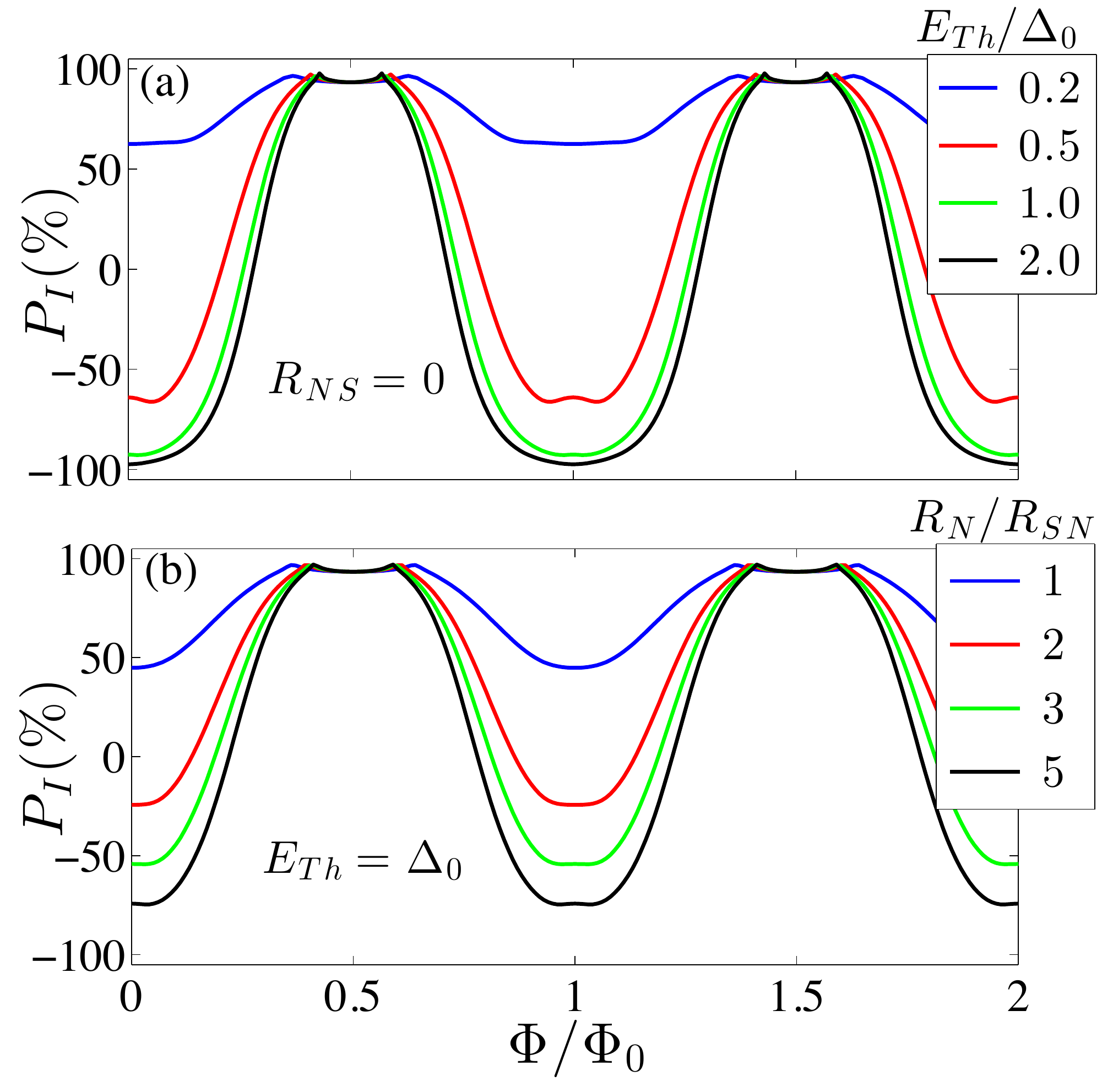}\vspace{-2mm}
\caption{Phase dependence of the current spin-polarization $P_I$  in the case   of a highly-transparent SN interface and different lengths of the wire, (a), and for $E_{Th}=\Delta_0$ and different values of the interface resistance, (b). In all calculations we set  $eV=h=0.4\Delta_0$. \label{fig4}}
\end{figure}
A finite interface resistance $R_{SN}$ at the SN contact has a similar effect on the magnitude and modulation amplitude of the spin-polarization. 
According to Eq. (\ref{BC}) the strength of the proximity effect is related to the coefficient $\gamma$. 
For a highly-transparent SN interface ($\gamma\rightarrow \infty$) the proximity effect in the wire is maximized, and in turn the modulation of the DoS in the metal.  
For finite values of $R_{SN}$, the suppression of the proximity effect leads to a weaker dependence of the DoS on the magnetic flux and, accordingly, the switching effect is suppressed. 
This is shown in  Fig. 4(b), where the $P_I(\Phi)$ dependence is plotted for different values of  $\gamma=R_N/R_{SN}$ at $E_{Th}=\Delta_0$ and $eV=h=0.4\Delta_0$. 
It can be seen that by decreasing $\gamma$ the switching occurs within a smaller range of $P_I$. 
We note again, that  even for $R_{SN}$ of the order of $\sim R_N$, large $P_I$ values can be achieved as well for proper applied magnetic fluxes [see the blue curve in Fig. 4(b)].

The switching of $P_I$ is therefore optimized by having highly-transparent SN interfaces, and for intermediate-length or short N wires (i.e., $E_{Th}\gtrsim \Delta_0$). 
In general, tuning of the phase bias in the structure can be achieved experimentally through an integrated  superconducting coil providing a suitable magnetic flux which allows, in principle, high-frequency   
operation.
In this context, the characteristic polarization switching frequency is given by $f={\rm min}[ E_{Th}/(2\pi \hbar),1/(2\pi\sqrt{LC}),\Delta_0/(2\pi \hbar)]$, i.e., it 
 is determined by the minimum among the inverse time that the DoS requires to follow a change in the phase difference across the N wire, the characteristic frequency of the LC phase biasing circuit
(where $L$ denotes the inductance  and $C$ is the total capacitance), and the characteristic frequency of the superconductor.
As $E_{Th}/(2\pi \hbar)\sim 10^{10}$ Hz for intermediate-length N wires,  $f$ can therefore easily approach values as high as $\sim 10^{10}$ Hz
for suitable $L$ and $C$  parameters.
By contrast, switching the spin polarization by changing the voltage bias across the tunnel junction is normally much slower. In such a case, $f$ can be estimated to be of the order of $\sim 1/(2\pi R_tC)$ therefore yielding at most $\sim 10^3\ldots 10^4$ Hz as the relevant switching frequency achievable in a typical cryogenic setup.  

We finally discuss two  conditions required for a correct operation of the spin-current injector:  (i) The avoidance of magnetic hysteresis, and (ii)  the occurrence of a good phase biasing in the structure. The first condition imposes that $2\pi I_J\mathcal{L}_G\lesssim \Phi_0$ \cite{Tinkham}, where $I_J$ is the Josephson supercurrent circulating along the loop, and  $\mathcal{L}_G$ is the ring geometric inductance.  Condition (ii) ensures that the phase difference set by the magnetic flux drops entirely at the wire ends, allowing a full modulation of its DoS. This condition can 
 be expressed as $\mathcal{L}_K^{ring}\ll \mathcal{L}_K^{N}$ \cite{Meschke,Saclay}, where $\mathcal{L}_K^{ring}\simeq \hbar R_{ring}/\pi \Delta_0$ is the ring kinetic inductance \cite{Tinkham} and $R_{ring}$ is the loop normal-state resistance, while $\mathcal{L}_K^{N}\simeq \hbar R_{N}/\pi \Delta_g$ is the wire kinetic inductance. Experiments have shown that  both conditions can be fulfilled by a proper choice of  materials and a suitable geometry \cite{Meschke,Saclay}.

{\it CONCLUSIONS.-} In conclusion, we  have proposed a hybrid quantum interference nanodevice that can be used as an efficient spin-current injector with controllable degree of  current polarization. 
The device operates by combining  phase-dependent superconducting proximity effect and an effective  Zeemann splitting of the density of states  induced by a ferromagnetic layer.
Under optimal conditions it  can provide strongly polarized (i.e., up to $\sim 100\%$)  spin-up and spin-down  currents in two ways: either by tuning an externally-applied magnetic flux or by changing the voltage bias across the structure.  In the former case,  switching frequencies of the order of tens of GHz can be achieved. Conventional metals combined  with ferromagnetic insulators such as, for instance, Eu chalcogenides layers \cite{Moodera1990,Santos,Miao} appear as promising materials for the implementation of this proposal.


{\it ACKNOWLEDGEMENTS.-}  F.G. acknowledges the FP7 program
No. 228464 MICROKELVIN, the Italian Ministry of Defense through the PNRM project TERASUPER, and the
Marie Curie Initial Training Action (ITN) Q-NET 264034 for partial financial support.
The work of F.S.B  was supported by the Spanish Ministry of Economy
and Competitiveness under Project FIS2011-28851-C02-02. F.S.B thanks Prof. Martin
Holthaus and his group for their kind hospitality at the Physics Institute of the 
Oldenburg University.

\end{document}